\documentclass[smallabstract,smallcaptions]{dccpaper}

\usepackage{tikz}
\usetikzlibrary{shapes}
\usepackage{pgfplots}
\pgfplotsset{compat=1.13}
\usetikzlibrary{shapes,arrows}
\usetikzlibrary{positioning}
\usepgfplotslibrary{colorbrewer}
\pgfplotsset{cycle list/Set2}
\usepackage[font={small,it}]{caption}

\usepackage{pgf}
\usepackage{varwidth}
\usetikzlibrary{calc} 

\usepackage[nolist]{acronym}

\usepackage{epsfig}
\usepackage{amsmath}
\usepackage{amssymb}
\usepackage{color}
\usepackage{url}
\usepackage{enumitem}

\usepackage{colortbl}

\usepackage{enumitem}
\usepackage{subfig}
\usepackage{placeins}

\usepackage[american]{babel}

\newlength{\figurewidth}
\newlength{\smallfigurewidth}

\usetikzlibrary{arrows, decorations.markings}
\usetikzlibrary{shapes}

\newcommand {\figref}[1]{Fig.~\ref{fig:#1}}

\newcommand {\eg}{\emph{e.g.}}
\newcommand {\ie}{\emph{i.e.}}

\newcommand{\colorND}{blue}
\newcommand{\colorNU}{lime}
\newcommand{\colorQuotients}{teal}
\newcommand{\colorU}{violet}
\newcommand{\colorReminders}{red}

\newdimen\zerolinewidth
\tikzset{
  zero line width/.code={%
    \zerolinewidth=\pgflinewidth
    \tikzset{line width=0cm}%
  },
  use line width/.code={%
    \tikzset{line width=\the\zerolinewidth}%
  },
  draw with anchor in boundary/.style={
    zero line width,
    postaction={draw,use line width},
  },
}

\newcolumntype{H}{>{\centering\arraybackslash}p{2em}}

\begin{acronym}
\acro{VF}{Variable-to-Fixed}
\acro{FV}{Fixed-to-Variable}
\acro{HT}{High Throughput}
\acro{ANS}{Asymmetric Numeral Systems}
\acro{TLB}{Translation Lookaside Buffer}
\acro{NCB}{Non-Compressible Bit}
\end{acronym}

\newcommand\blfootnote[1]{%
  \begingroup
  \renewcommand\thefootnote{}\footnote{#1}%
  \addtocounter{footnote}{-1}%
  \endgroup
}

\begin{document}

\title
{\large
\textbf{Rice-Marlin Codes: \\ Tiny and Efficient Variable-to-Fixed Codes}
}

\author{%
{\small\begin{minipage}{\linewidth}\begin{center}
Manuel Martinez$^{\ast}$ and Joan Serra-Sagrist\`{a}$^{\dag}$\\[0.75em]
\begin{tabular}{ccc}
$^{\ast}$Karlsruhe Institute of Technology & \hspace*{0.0in} & $^{\dag}$Universitat Aut\`{o}noma de Barcelona \\
Karlsruhe, 76131, Germany && Cerdanyola del Vall\`{e}s, 08193, Spain\\
\url{manuel.martinez@kit.edu} && \url{Joan.Serra@uab.cat}
\end{tabular}
\end{center}\end{minipage}}
}

\maketitle
\thispagestyle{empty}

\begin{abstract}
Marlin~\cite{martinez2017marlin,martinez2018improving}\blfootnote{
\footnotesize
JSS acknowledges support from Spanish Ministry of Economy and Competitiveness (MI\-NE\-CO) and European Regional Development Fund (FEDER) under Grant TIN2015-71126-R, and from Catalan Government under Grant 2017SGR-463.
This work is also partially supported by the German Federal Ministry of Education and Research (BMBF) within the KonsensOP project.
} is a \acf{VF} codec optimized for high decoding speed through the use of small sized dictionaries that fit in the L1 cache of most CPUs.
While the size of Marlin dictionaries is adequate for decoding, they are still too large to be encoded fast.
We address this problem by proposing two techniques to reduce the alphabet size.
The first technique is to encode rare symbols in their own segment, and the second is to combine Marlin dictionaries with Rice encoding, hence our name Rice-Marlin for our new codec.
Using those techniques, we are able to reduce the size of Marlin dictionaries by a factor of 16, 
not only enabling faster encoding speed, but also achieving better compression efficiency.
\end{abstract}

\section{Introduction}

\ac{HT} codecs prime speed over compression efficiency.
While \ac{HT} codecs have been used to reduce storage needs (\ie, LZ4~\cite{LZ4}, and LZO~\cite{LZO}), their main application is to improve the throughput of communication interfaces~\cite{SNAPPY}.


Most generic lossless compression codecs that are not considered to be \ac{HT} are based on the deflate~\cite{deflate} algorithm, which is a two-step algorithm that combines dictionary compression(\ie, LZ77~\cite{ziv1977universal}), and an entropy encoder (\ie, Huffman~\cite{huffman1952method}, arithmetic~\cite{rissanen1976generalized} or range encoding~\cite{martin1979range}).
However, entropy encoding is significantly slower than the dictionary compression step and, furthermore, in most target applications, entropy coding provides reduced compression gains as compared to those due to dictionary compression.
Hence, in most \ac{HT} algorithms, the entropy encoding step is either dropped~\cite{LZ4,SNAPPY} or severely simplified~\cite{LZO,lenhardt2012gipfeli}.
In particular, most research in \ac{HT} focuses on finding novel ways to find matches in LZ77 dictionaries that offer compelling tradeoffs between coding speed and efficiency~\cite{fastLZ,harnik2014fast}.



Still, there are applications that need a fast entropy codec.
In particular data obtained by sensors (\ie, image, audio, etc.) generally can not be compressed well using generic dictionary approaches; hence, LZ77 based \ac{HT} codecs do not perform well in this kind of data.
One option to deal with this problem is to use fast versions of generic entropy codecs (\ie, Huff0~\cite{Huff0}
for Huffman and FSE~\cite{FSE} for \acl{ANS}~\cite{DBLP:journals/corr/abs-0902-0271}).
However, if we know that the source we are compressing provides symbols following a particular probability distribution, we can use a codec that specifically matches this distribution.
The typical example of this case is when encoding sensor residuals, which often follow Laplacian distributions. 
For this particular distribution, the Rice codec~\cite{Rice} is close to optimal, and faster than generic entropy codecs.
Hence, Rice encoding is very popular in lossless image codecs~\cite{Lagarith,CharLS}.

Rice codec is a \ac{FV} codec, where each input symbol is encoded as a variable number of bits. 
Generally, \ac{FV} codecs can be encoded exceedingly fast using a lookup table, but are slower to decode as the boundaries between compressed symbols are not known in advance.
On the other hand, \acf{VF}  codecs (\eg, Tunstall~\cite{tunstall1967synthesis}) can be decompressed fast, but are slower to compress.

Marlin~\cite{martinez2017marlin,martinez2018improving} was the first \ac{VF} codec that achieved a competitive compression ratio at \ac{HT} decoding speeds thanks to the combination of using plurally parsable dictionaries~\cite{savari1999variable,al2015using,yoshida2010efficient,yamamoto2001average,dube2018individually} in a memoryless context, however, being a \ac{VF} codec, Marlin is slow to encode.
The main bottleneck during the encoding process is related to the size of the dictionary.
For decoding, we only need to store the dictionary itself on the L1 cache, but for encoding, we need to encode either a tree or a hash table whose size is several times larger than the dictionary itself.

In \ac{VF} codecs, the size of the dictionary must be larger than the alphabet size.
Hence, when encoding 8-bit sources the size of the dictionary must be larger than 256 entries.
As the next efficient value for the codeword size is 12-bit, Marlin dictionaries have 4096 entries.
This size poses two problems: first, the encoding tree does not fit in the L1 cache, and also 4096 entries are not enough for efficient encoding of high entropy sources.
We deal with those two problems in this contribution.

Here we apply two techniques that reduce the number of symbols to be encoded by Marlin, effectively decreasing the size of the Marlin alphabet as much as possible.

Our first technique is to remove non-compressible bits (\ie, bits whose probability of being 0 and 1 is close to 50\%, and are uncorrelated to other bits).
Having non-compressible bits in an alphabet has an extremely negative effect on its efficiency, and it is thus worth just to store them in a binary form, similarly to what Rice codes do.
Hence we named the combination Rice-Marlin codec. 
Each stored bit halves the alphabet size, duplicating the ratio between dictionary size and alphabet size.

Our second technique consists of removing from the alphabet those symbols whose probability of being emitted in a message is below a certain threshold, becoming then unrepresented symbols.
Each occurrence of an unrepresented symbol is stored independently in a dedicated section.

The first technique is more effective when compressing high entropy sources, while the second technique is more effective when compressing low entropy sources.
By combining both techniques we are able to use 8-bit codewords along all the entropy ranges, while improving the compression ratio over previous approaches.

Experimental results for both synthetic and real data reveal how Rice-Marlin codes achieve better compression ratio, the same decoding speed, and 2.4x faster encoding speed when compared to original Marlin codes.

\clearpage
\section{Background: Marlin Dictionaries}

In this section we review the concepts from Marlin dictionaries that are necessary to present our new contributions.

\begin{figure}[ht!]
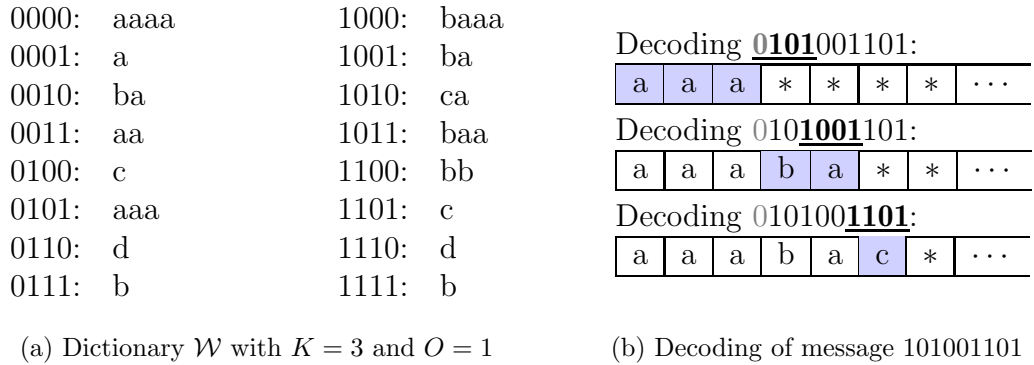

\centering
\subfloat[Dictionary $\mathcal{W}$ with $K=3$ and $O=1$]{
\begin{minipage}[b][4.5cm][c]{.25\textwidth}
\centering
	\begin{tabular}{r l}
	0000: & aaaa \\
	0001: & a \\
	0010: & ba \\
	0011: & aa \\
	0100: & c \\
	0101: & aaa \\
	0110: & d \\
	0111: & b 
	\end{tabular}
\end{minipage}
\quad
\begin{minipage}[b][4.5cm][c]{.25\textwidth}
\centering
	\begin{tabular}{r l}
	1000: & baaa \\
	1001: & ba \\
	1010: & ca \\
	1011: & baa \\
	1100: & bb \\
	1101: & c \\
	1110: & d \\
	1111: & b 
	\end{tabular}
\end{minipage}
}
\quad
\subfloat[Decoding of message 101001101]{
\begin{minipage}[b][4.5cm][c]{.35\textwidth}
Decoding {\fontseries{b}\selectfont \underline{{\color{gray}0}101}}001101:

	\begin{tabular}{|c|c|c|c|c|c|c|c}
	\hline
	\cellcolor{blue!18}{a} & \cellcolor{blue!18}{a} & \cellcolor{blue!18}{a} & $\ast$ & $\ast$ & $\ast$ & $\ast$ & $\cdots$ \\
	\hline
	\end{tabular}

\vspace{.125cm}
Decoding {\color{gray}0}10{\fontseries{b}\selectfont\underline{1001}}101:

	\begin{tabular}{|c|c|c|c|c|c|c|c}
	\hline
	{a} & {a} & {a} & \cellcolor{blue!18}{b} & \cellcolor{blue!18}{a} & $\ast$ & $\ast$ & $\cdots$ \\
	\hline
	\end{tabular}

\vspace{.125cm}
Decoding {\color{gray}0}10100{\fontseries{b}\selectfont\underline{1101}}:

	\begin{tabular}{|c|c|c|c|c|c|c|c}
	\hline
	a & a & a & b & a & \cellcolor{blue!18}{c} & $\ast$ & $\cdots$ \\
	\hline
	\end{tabular}
\end{minipage}
}
\caption{
Decoding example for the message 101001101 using the dictionary $\mathcal{W}$.
Codewords of $\mathcal{W}$ are encoded in 4 bits, but only 3 bits are consumed from the compressed source at each step ($K=3$), and there is 1 bit of overlap between consecutive codewords ($O=1$).
The grayed bit corresponds to the initialization which in our current algorithm is always zero.
}
\label{fig:example}
\end{figure}

Marlin expects an alphabet $A$ that contains a certain number of symbols and their probabilities, and builds a dictionary $\mathcal{W}$ with $2^N$ words.
Most \ac{VF} codecs do uncompress a message by consuming $N$ bits from the compressed stream and emit its corresponding word.
Marlin uses instead an overlapping codeword methodology~\cite{martinez2018improving}, where $N$ bits are \emph{peeked} from the stream, corresponding to one entry in the dictionary, but only $K$ bits are consumed, as seen in \figref{example}.
In our notation, $K$ is the number of bits that are consumed from the source at each iteration, and $O$ is the number of bits that overlap between codewords, with $N = O + K$.


Because of the overlap, not all possible words of the dictionary are accessible at each step.
This is due to the prefix which is fixed by the previous codeword.
We named \emph{chapters} to each set of accessible words.
On the example represented in \figref{example}, we have two chapters.
The first contains the codewords from $0000$ to $0111$, and the second chapter contains the codewords from $1000$ to $1111$.
There are not repeated words within a chapter, but the same word can appear in different chapters.

\acp{VF} codes, including Marlin, have the limitation that the number of words in the dictionary $|\mathcal{W}|$ must be larger than the number of symbols of the alphabet $|A|$.
Marlin requisites are more strict due to the overlapping, and  $2^K$ must be larger than $|A|$ to achieve compression.
Hence, in previous publications, we have suggested to use a value 12 for $K$ when encoding 8-bit alphabets, forcing us upon dictionaries that contain thousands of words.
This limit is the motivation for the Rice-Marlin codecs introduced in this work, which allow to reduce the size of the working alphabet and, therefore, to use a value of 8 for $K$.




\clearpage
\section{Increasing the Efficiency of Marlin Dictionaries}

The coding efficiency of stateless \ac{VF} codecs is related to the ratio between the size of the dictionary and the size of the alphabet.
However, increasing the size of the dictionary requires more memory and reduces the speed of the codec.
Therefore, in this section we present two techniques that allow us to increase the efficiency of \ac{VF} codecs by reducing the size of the alphabet instead of increasing the size of the dictionary.
The first technique, \emph{Removing High Entropy Bits}, is more effective on high entropy sources, while the second technique, \emph{Removing Low Entropy Symbols}, is more effective on low entropy sources.



\subsection{Removing High Entropy Bits}

We analyze source symbols as a group of bits. 
Often, when encoding high entropy sources, the least significant order bits of each symbol can not be compressed (\ie, their probability of being 0 or 1 is close to 50\% and uncorrelated with other bits).
This is a well known effect, and it is taken advantage from in, \eg, Rice~\cite{Rice} codecs, where the least significant order bits are simply stored in a truncated binary form.

In \ac{VF} codecs, each non-compressible bit effectively duplicates the number of alphabet symbols, 
hence, by encoding them in a separate section, we free a large number of codewords in the dictionary.

We use a simple variation of the Rice-Golomb coding as we split each input symbol in two parts:
the S least significant bits are the \emph{reminder}, while the most significant bits are the \emph{quotient}, defined as:
\begin{equation}
r = x\text{ mod } 2^S , \text{ and } q = \left \lfloor \frac{x}{2^S} \right \rfloor \, , 
\label{eq:eqw}
\end{equation}
where $x$ is the original symbol interpreted as an unsigned integer.

For each input message, we compress all \emph{quotients} together using the Marlin codec, 
while the reminders are simply packed together afterwards.
Hence, for $S=0$, our codec is simply equivalent to the original Marlin code.

To find the optimal $S$ for an alphabet, we build a dictionary for each possible value of $S$ and choose the one with the best efficiency.

\subsection{Removing Low Entropy Symbols}

While building a \ac{VF} codec, we must ensure that each possible source symbol can be represented.
As a consequence, symbols that are unlikely to be generated by the source must use at least one entry in the dictionary.
When using very small dictionaries together with low entropy sources, a significant portion of the entries in the dictionary are wasted representing symbols whose probability of appearing in a message is close to zero.

To solve this problem, we identify symbols whose occurrence probability is below a certain threshold, and we exclude them from the Marlin dictionary, labeling them as \emph{unrepresented symbols}.
When encoding a message, each \emph{unrepresented symbol} is stored uncompressed in a dedicated section.


\clearpage
\section{Proposed Code Format}

\begin{figure}[t!]

\tikzstyle{vecArrow} = [thick, decoration={markings,mark=at position
   1 with {\arrow[semithick]{open triangle 60}}},
   double distance=1.4pt, shorten >= 5.5pt,
   preaction = {decorate},
   postaction = {draw,line width=1.4pt, white,shorten >= 4.5pt}]

\tikzstyle{vecArrowRev} = [thick, decoration={markings,mark=at position
   0 with {\arrowreversed[semithick]{open triangle 60}}},
   double distance=1.4pt, shorten <= 5.5pt,
   preaction = {decorate},
   postaction = {draw,line width=1.4pt, white,shorten <= 4.5pt}]

\tikzstyle{innerWhite} = [semithick, white,line width=1.4pt, shorten >= 4.5pt]

\centering
\begin{tikzpicture}[thick]
  \node[draw,rectangle,minimum height=1.8cm,align=center] (encoder) {Marlin\\ Encoder};
  \node[draw,rectangle,left of=encoder,node distance=3cm,align=center] (block) { Source Block };
  \node[draw,fill=white, text=white, rectangle, above of=block         , node distance=  1.25cm,align=center,minimum height=2em] (dictA) { Dictionaries };
  \node[draw,fill=white, text=white, rectangle, above left of=dictA, node distance=1.0em,align=center,minimum height=2em] (dictB) { Dictionaries };
  \node[draw,fill=white, text=white, rectangle, above left of=dictA, node distance=.75em,align=center,minimum height=2em] (dictC) { Dictionaries };
  \node[draw,fill=white, text=white, rectangle, above left of=dictA, node distance=.50em,align=center,minimum height=2em] (dictD) { Dictionaries };
  \node[draw,fill=white, text=white, rectangle, above left of=dictA, node distance=.25em,align=center,minimum height=2em] (dictE) { Dictionaries };
  \node[draw,fill=white, text=black, rectangle, above left of=dictA, node distance= .0em,align=center,minimum height=2em] (dict) { Dictionaries };

  \node[draw,rectangle,right of=encoder,node distance=2.8cm,align=center] (encoded) { Encoded \\ Message };

  \node[draw,rectangle,right of=encoded,node distance=2.8cm,minimum height=1.8cm,align=center] (decoder) {Marlin\\ Decoder};

  \node[draw,rectangle,right of=decoder,node distance=3cm,align=center] (oblock) { Output Block };

  \draw[vecArrow] (block.east) -- (encoder);
  \draw[vecArrow] (dict.east) -| (encoder);
  \draw[vecArrow] (dict.east) -| (decoder);
  \draw[innerWhite] (dict.east) -| (encoder);

  \draw[vecArrowRev] (encoded.west) -- ++(-0.98,0);

  \draw[vecArrow] (encoded.east) -- ++(+.98,0);

  \draw[vecArrow] (decoder) -- (oblock);

\end{tikzpicture}
\caption{
To improve coding speed, a set of dictionaries is built beforehand and known to both encoder and decoder. 
The encoder selects the best fitting dictionary, compresses the block, and emits the compressed message which includes the index of the dictionary used.
The decoder recovers the original data using the selected dictionary.
}
\vspace{-0.30cm}
\label{fig:architecture_overview}
\end{figure}
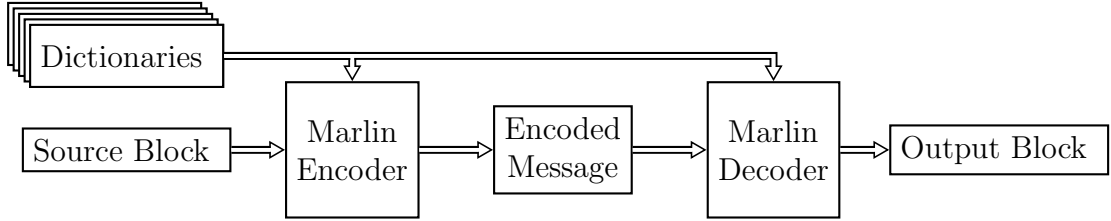

Most entropy encoders build custom encoding and decoding tables based on the statistics of the symbols appearing in the block being currently encoded. 
However, building such tables is time consuming, and several compression formats 
default to a predefined generic table, as in JPEG or ZIP.
Marlin is optimized for speed, but we also want it to provide a competitive compression ratio, 
hence, we use a hybrid approach where we build beforehand a set of dictionaries that cover a wide range of possible probability distributions that may appear in the source,
and we make this set of dictionaries available to both encoder and decoder, as seen in \figref{architecture_overview}.
Then, when compressing a block, we select the best fitting dictionary. 

Also, as in previous Marlin codes, we rely on the software used to transmit the message from source to target to provide the decoder with the size of the encoded message and the size of the original message.

Our suggested message format has the following sections: 
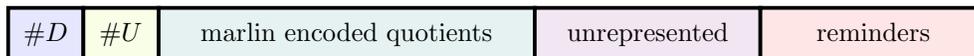
\begin{center}
\begin{footnotesize}
\begin{tikzpicture}[line width=0.5mm, minimum height=0.70cm]
  \node [draw with anchor in boundary, minimum width=1cm, fill={\colorND!10}] (nd) {$\#D$};
  \node [draw with anchor in boundary, minimum width=1cm, fill={\colorNU!10},right=0cm of nd] (nr) {$\#U$};
  \node [draw with anchor in boundary, minimum width=5cm, fill={\colorQuotients!10},right=0cm of nr] (marlin) {marlin encoded quotients};
  \node [draw with anchor in boundary, minimum width=3cm, fill={\colorU!10},right=0cm of marlin] (rare) {unrepresented};
  \node [draw with anchor in boundary, minimum width=3cm, fill={\colorReminders!10},right=0cm of rare] (reminders) {{\color{\colorReminders!10}p}reminders{\color{\colorReminders!10}p}};
\end{tikzpicture}
\end{footnotesize}
\end{center}
\begin{enumerate}[noitemsep,nolistsep]
  \item \emph{$\#D$:} 
1 byte containing the index of the dictionary used to encode the message.
  \item \emph{$\#U$:} 
1 byte containing the number of unrepresented symbols in the message.
  \item \emph{quotients:}
variable sized field encoding the quotients using a Marlin dictionary.
  \item \emph{unrepresented:}
each unrepresented symbol encoded as a pair $\{$\emph{location},\emph{symbol}$\}$.
  \item \emph{reminders:}
bit field with all concatenated reminders.
\end{enumerate}

The \emph{quotients} section is made of concatenated Marlin words, whose size depends on the chosen dictionary. 
The \emph{unrepresented} section is made by pairs $\{$\emph{location},\emph{symbol}$\}$,
those encode the symbols that the selected Marlin dictionary can not represent.
In most cases this section is small or even empty.
The \emph{location} field contains the absolute position of the \emph{symbol} within the uncompressed message, and its size depends precisely on the size of the uncompressed message: 1 byte for messages smaller than $2^8$ symbols, 2 bytes for message smaller than $2^{16}$ symbols, 4 bytes for messages smaller than $2^{32}$ symbols, etc.
If the original message has more than 255 \emph{unrepresented} symbols we store the original message uncompressed.

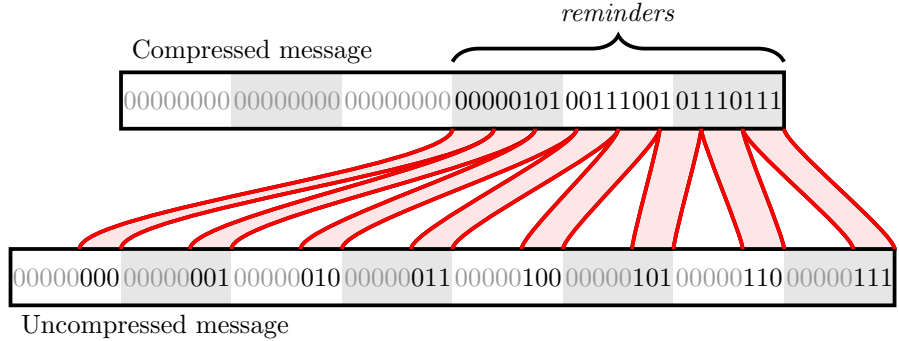
\begin{figure}[ht!]
\centering
\begin{footnotesize}
\begin{tikzpicture}[line width=0.5mm,  minimum width=2.0cm, minimum height=0.75cm]

\def\FillZone(#1,#2,#3,#4,#5,#6,#7){
  \filldraw[fill=#1!10] 
	($(#2.south west)!{#3}!(#2.south east)$) to [out=-90,in=90,looseness=0.25] ($(#5.north west)!{#6}!(#5.north east)$) --
	($(#5.north west)!{#7}!(#5.north east)$) to [out=90,in=-90,looseness=0.25] ($(#2.south west)!{#4}!(#2.south east)$) -- cycle;

  \draw [-,#1] ($(#2.south west)!{#3}!(#2.south east)$) to [out=-90,in=90,looseness=0.25]   ($(#5.north west)!{#6}!(#5.north east)$);
  \draw [-,#1] ($(#5.north west)!{#7}!(#5.north east)$) to [out=90,in=-90,looseness=0.25] ($(#2.south west)!{#4}!(#2.south east)$);
  
};

\def\Decorate(#1,#2,#3,#4){

  \draw [decorate,decoration={brace,amplitude=10pt,raise=4pt},yshift=0pt] ($(#1.north west)!{#2}!(#1.north east)$) -- ($(#1.north west)!{#3}!(#1.north east)$) node [black,midway,yshift=0.8cm] {\footnotesize \emph{#4}};

};

 \node [
 draw with anchor in boundary,
 anchor=north,
 rectangle split,
 rectangle split parts=6, 
 rectangle split horizontal,
 rectangle split part fill={white,black!10,white,black!10,white,black!10,white,black!10},
 rectangle split draw splits=false,
 inner sep=0.9pt]
 (compressed) {
\nodepart{one}{\color{black!40}00000000}
\nodepart{two}{\color{black!40}00000000}
\nodepart{three}{\color{black!40}00000000}
\nodepart{four}{00000101}
\nodepart{five}{00111001}
\nodepart{six}{01110111}
};

\node [
 above=0.65cm of compressed.west,
 anchor=west] (label_compressed) {Compressed message};

\Decorate(compressed,8/16,16/16,reminders);

 \node [
 draw with anchor in boundary,
 below=1.6cm of compressed,
 anchor=north,
 rectangle split,
  rectangle split parts=8, 
 rectangle split horizontal,
 rectangle split part fill={white,black!10,white,black!10,white,black!10,white,black!10,white},
 rectangle split draw splits=false,
 inner sep=0.9pt]
 (uncompressed) {
\nodepart{one}{{\color{black!40}00000}000}
\nodepart{two}{{\color{black!40}00000}001}
\nodepart{three}{{\color{black!40}00000}010}
\nodepart{four}{{\color{black!40}00000}011}
\nodepart{five}{{\color{black!40}00000}100}
\nodepart{six}{{\color{black!40}00000}101}
\nodepart{seven}{{\color{black!40}00000}110}
\nodepart{eight}{{\color{black!40}00000}111}
};


\FillZone(red,compressed,(3+0/8*3)/6,(3+(1/8)*3)/6,uncompressed,0/8+(5/(8*8)),0/8+(8/(8*8)));
\FillZone(red,compressed,(3+1/8*3)/6,(3+(2/8)*3)/6,uncompressed,1/8+(5/(8*8)),1/8+(8/(8*8)));
\FillZone(red,compressed,(3+2/8*3)/6,(3+(3/8)*3)/6,uncompressed,2/8+(5/(8*8)),2/8+(8/(8*8)));
\FillZone(red,compressed,(3+3/8*3)/6,(3+(4/8)*3)/6,uncompressed,3/8+(5/(8*8)),3/8+(8/(8*8)));
\FillZone(red,compressed,(3+4/8*3)/6,(3+(5/8)*3)/6,uncompressed,4/8+(5/(8*8)),4/8+(8/(8*8)));
\FillZone(red,compressed,(3+5/8*3)/6,(3+(6/8)*3)/6,uncompressed,5/8+(5/(8*8)),5/8+(8/(8*8)));
\FillZone(red,compressed,(3+6/8*3)/6,(3+(7/8)*3)/6,uncompressed,6/8+(5/(8*8)),6/8+(8/(8*8)));
\FillZone(red,compressed,(3+7/8*3)/6,(3+(8/8)*3)/6,uncompressed,7/8+(5/(8*8)),7/8+(8/(8*8)));

\node [
 below=0.65cm of uncompressed.west,
 anchor=west] (label_uncompressed) {Uncompressed message};

\end{tikzpicture}
\end{footnotesize}
\caption{
Reminders are stored concatenated all together in its own section.
Both encoding and encoding of reminders can be done very efficiently using bit manipulation instructions.
}
\vspace{-0.30cm}
\label{fig:reminders}
\end{figure}

The \emph{reminders} section is made of the concatenated least significant bits from the original message, as seen in \figref{reminders}.
Those can be encoded and decoded fast using bit shuffling instructions.

\section{Improved Encoding Algorithm}
\label{ch:implementation}

The process of encoding of a Marlin message is divided in three stages.
On the first stage, we calculate the quotients for each symbol and we encode them using the \ac{VF} dictionary.
In this step, symbols whose quotient is not represented in the \ac{VF} alphabet are replaced by the most common quotient acting as a placeholder.
The second stage is to emit the location and value of each \emph{unrepresented} symbol identified in the previous step.
The third stage is to emit the \emph{reminders} of each source symbol.

The most time consuming stage is, by far, the encoding of the \ac{VF} words.
The goal of this stage is to find the largest prefix of the source that corresponds to a word in the dictionary.
The two common approaches for solve this problem (hash tables or prefix trees) require one query to the data structure per source symbol;
as this query can not be predicted, it often causes a cache-miss, becoming the main bottleneck.
Hence, to speed up encoding times we must have better locality to improve the cache hits/miss ratio.
We found out that prefix trees enable us to better control locality.


We build our prefix tree in the form of a matrix organized as shown in \figref{encoder}.
Each column of the encoding matrix represents the current state, and each row represents the next source symbol to be encoded.
Each codeword in the dictionary corresponds to one state, and each cell contains both the next state, and a flag that indicates if the current state/codeword must be emitted in the compressing stream.
For example, if we have encoded \emph{"aa"} and thus we are on the fourth column, and we fetch an \emph{"a"} from the source, we must fetch the first row of the fourth column, which would indicate to go to the state belonging to the \emph{"aaa"}, and not emit a word.
On the other hand, if we are in \emph{"aa"} and receive a \emph{"b"}, we fetch the second row of the fourth column, and thus we will emit the codeword corresponding to \emph{"aa"} (\ie, the column index), and go to the state \emph{"b"} of the corresponding chapter.

This data structure has three particularities:
First, we avoid one memory fetch by identifying codeword with state.
Second, we avoid one memory fetch by storing the next state and the flag in the same cell.
Third, by choosing this disposition of rows and columns, and sorting inputs symbols from most probable to least probable, we will create an inbalance where most memory fetches will be close together and centered on the first rows of the structure, as seen in \figref{encoder}, improving locality.

\begin{figure}[ht!]
\begin{footnotesize}
\begin{center}
\begin{tabular}{ r *{8}{|c}||c*{7}{|c}}
     & {\color{red}aaaa}& {\color{red}a}& {\color{red}ba}& {\color{red}aa}& {\color{red}c}& {\color{red}aaa}& {\color{red}d}& {\color{red}b}& {\color{blue}baaa}& {\color{blue}ba}& {\color{blue}ca}& {\color{blue}baa}& {\color{blue}bb}& {\color{blue}c}& {\color{blue}d}& {\color{blue}b}\\ \hline
  a  & \cellcolor{black!70}{\color{red}a!} & \cellcolor{black!78}{\color{red}aa} & \cellcolor{black!53}{\color{red}a!} & \cellcolor{black!52}{\color{red}aaa} & \cellcolor{black!27}{\color{red}a!} & \cellcolor{black!34}{\color{red}aaaa} & \cellcolor{black!9}{\color{red}a!} & \cellcolor{black!26}{\color{red}ba} & \cellcolor{black!60}{\color{red}a!} & \cellcolor{black!44}{\color{blue}baa} & \cellcolor{black!56}{\color{red}a!} & \cellcolor{black!29}{\color{blue}baaa} & \cellcolor{black!46}{\color{red}a!} & \cellcolor{black!28}{\color{blue}ca} & \cellcolor{black!28}{\color{red}a!} & \cellcolor{black!21}{\color{blue}ba} \\ \hline
  b  & \cellcolor{black!23}{\color{red}b!} & \cellcolor{black!26}{\color{blue}b!} & \cellcolor{black!18}{\color{red}b!} & \cellcolor{black!17}{\color{blue}b!} & \cellcolor{black!9}{\color{red}b!} & \cellcolor{black!11}{\color{blue}b!} & \cellcolor{black!3}{\color{red}b!} & \cellcolor{black!9}{\color{blue}b!} & \cellcolor{black!20}{\color{red}b!} & \cellcolor{black!15}{\color{blue}b!} & \cellcolor{black!19}{\color{red}b!} & \cellcolor{black!10}{\color{blue}b!} & \cellcolor{black!15}{\color{red}b!} & \cellcolor{black!9}{\color{blue}b!} & \cellcolor{black!9}{\color{red}b!} & \cellcolor{black!7}{\color{blue}bb} \\ \hline
  c  & \cellcolor{black!8}{\color{red}c!} & \cellcolor{black!9}{\color{blue}c!} & \cellcolor{black!6}{\color{red}c!} & \cellcolor{black!6}{\color{blue}c!} & \cellcolor{black!3}{\color{red}c!} & \cellcolor{black!4}{\color{blue}c!} & \cellcolor{black!1}{\color{red}c!} & \cellcolor{black!3}{\color{blue}c!} & \cellcolor{black!7}{\color{red}c!} & \cellcolor{black!5}{\color{blue}c!} & \cellcolor{black!6}{\color{red}c!} & \cellcolor{black!3}{\color{blue}c!} & \cellcolor{black!5}{\color{red}c!} & \cellcolor{black!3}{\color{blue}c!} & \cellcolor{black!3}{\color{red}c!} & \cellcolor{black!2}{\color{blue}c!} \\ \hline
  d  & \cellcolor{black!2}{\color{red}d!} & \cellcolor{black!3}{\color{blue}d!} & \cellcolor{black!2}{\color{red}d!} & \cellcolor{black!2}{\color{blue}d!} & \cellcolor{black!1}{\color{red}d!} & \cellcolor{black!1}{\color{blue}d!} & \cellcolor{black!0}{\color{red}d!} & \cellcolor{black!1}{\color{blue}d!} & \cellcolor{black!2}{\color{red}d!} & \cellcolor{black!1}{\color{blue}d!} & \cellcolor{black!2}{\color{red}d!} & \cellcolor{black!1}{\color{blue}d!} & \cellcolor{black!1}{\color{red}d!} & \cellcolor{black!1}{\color{blue}d!} & \cellcolor{black!1}{\color{red}d!} & \cellcolor{black!0}{\color{blue}d!} \\ \hline
\end{tabular}
\end{center}
\end{footnotesize}
\vspace{-0.5cm}
\caption{
\ac{VF} encoding table for the dictionary shown in \figref{example}.
Our current state is encoded in a column, and corresponds to the current codeword.
The next possible symbols are encoded as rows.
Each cell contains our next state (our next codeword), as well as a flag (!) indicating that the current codeword should be emited.
Shading color indicates access probability. 
Red and blue symbols correspond to the two chapters of the dictionary.
}
\label{fig:encoder}
\end{figure}

\section{Evaluation}

We integrated our new contributions in Marlin's codebase, which is publicly available\footnote{Git repository: \url{https://github.com/MartinezTorres/marlin} (tag:\texttt{dcc2019})}.
Evaluation is performed on a i5-6600K CPU at 3.5GHz with 64GB of DDR4-2133 RAM running Ubuntu 16.04 and compiled using GCC v5.4.0 with the \texttt{-O3} flag.
While previous Marlin codes suggested to use a $K = 12$, our new contributions allow us to use a $K = 8$, avoiding shift operations in the decoding loop.
We use $O=4$ as the overlap parameter.
In this evaluation we decorate Marlin codes with the notation $(K,O)$ to indicate the used values for $K$ and $O$.
Our main evaluation metric is the \emph{compression efficiency} 
 defined as the ratio between the information entropy of the source and the average bit rate achieved: $\eta_X = H(X)/\text{ABR}(X)$.

\subsection{Results on synthetic data}

\begin{figure}[h!]
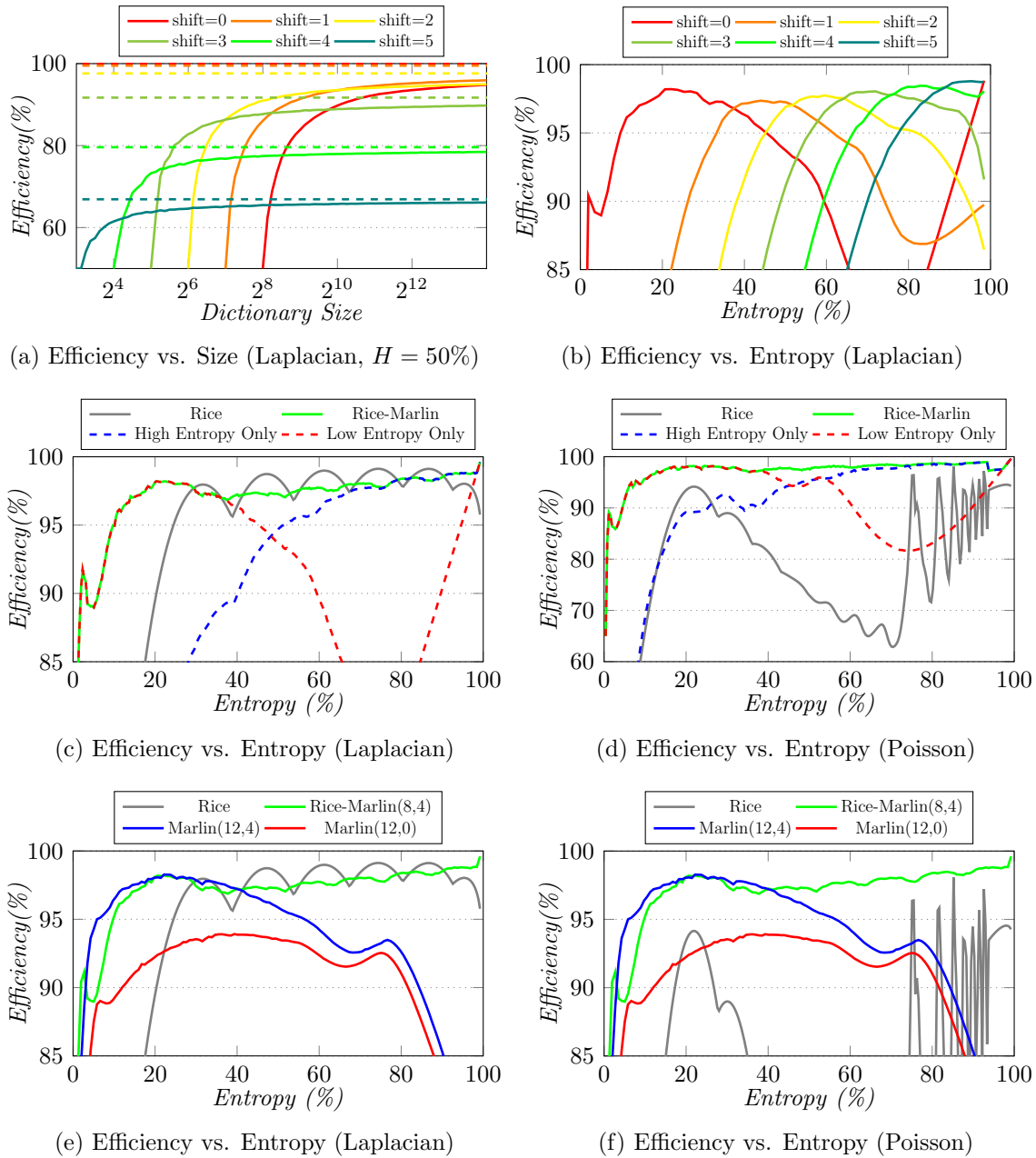

\centering
\subfloat[Efficiency vs. Size (Laplacian, $H=50\%$)] {

\label{fig:synA}
}
\subfloat[Efficiency vs. Entropy (Laplacian)] {
%
\label{fig:synB}
}

\subfloat[Efficiency vs. Entropy (Laplacian)] {
%
\label{fig:synC}
}
\subfloat[Efficiency vs. Entropy (Poisson)] {
%
\label{fig:synD}
}

\subfloat[Efficiency vs. Entropy (Laplacian)] {
%
\label{fig:synE}
}
\subfloat[Efficiency vs. Entropy (Poisson)] {
%
\label{fig:synF}
}
\caption{
(a): larger shift values enable smaller dictionary sizes, but reduce the maximum possible efficiency achievable, which we represent in a dashed line.
(b): as a consequence, we must find the right balance for each entropy level.
(c) and (d): removing low entropy symbols is complementary to removing high entropy bits, and combining both provides excellent efficiency over the entire entropy range.
Unlike Rice codec, our Rice-Marlin codec is generic and performs well even in non Laplacian/Exponential distributions.
(e) and (f): compared to previous versions of Marlin, Rice-Marlin achieves approximately the same compression efficiency for low entropy sources while using 16x smaller dictionaries, while having significantly better efficiency for high entropy sources.
}
\label{fig:synthetic}
\end{figure}

We have evaluated our new contributions on synthetic data distributions at different entropies.
In \figref{synA} we show how using larger shift ($S$) values in our Rice-Marlin distributions allow us to achieve compression gains with smaller dictionary sizes, but also reduces the maximum efficiency achievable, as the stored reminders are not compressed at all.
In \figref{synB} we observe how we must find the right shift value to use for each entropy level, 
with high entropy levels requiring larger shifts.

In \figref{synC} and \figref{synD} we have analyzed the impact of our high and low entropy contributions on different probability distributions, showing that both are complementary, and together we achieve compression efficiencies well above the 95\% mark in almost the entire range of entropies and tested distributions.

The Rice codec, being a specialized codec, is very efficient for Laplacian distributions but struggles on significantly different distributions, such as Poisson.
On the other hand, Marlin is a generic \ac{VF} codec that can efficiently compress sources from any probability distribution, even when used in combination with Rice encoding.

Finally, in \figref{synE} and \figref{synF}, we show our improvement against our previous versions of Marlin, which use dictionary sizes 16 times larger.
There we achieve approximately the same accuracy in low entropy sources, as the large dictionaries used previously were not very sensitive to the few codewords wasted on rare symbols.
However, we observe a significant improvement on high entropy sources.

\clearpage
\subsection{Results on real data}

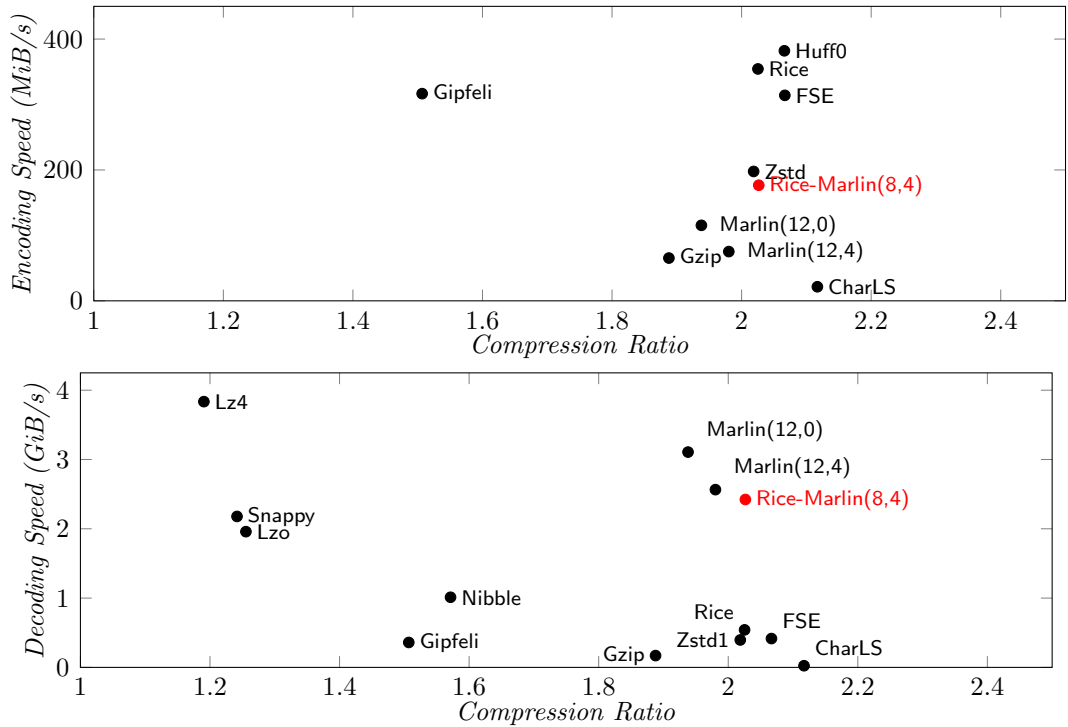
\begin{figure}[t]
\centering
\begin{tikzpicture}
\begin{axis}[
x tick label style={font={\footnotesize},yshift=0mm}, 
y tick label style={font={\footnotesize},xshift=0mm},
xlabel style={font={\footnotesize},yshift= 2mm}, 
ylabel style={font={\footnotesize},yshift=-2mm},
height=5.5cm, width=14.5cm, ymin=0, ymax=450, xmin=1, xmax=2.5, 
xlabel=\emph{Compression Ratio}, ylabel=\emph{Encoding Speed (MiB/s)}]
	\addplot[color=black,mark=*,only marks, point meta=explicit symbolic, nodes near coords,every node near coord/.append style={font={\sffamily\scriptsize},rotate=0, anchor=\myanchor},x=x,y=y,visualization depends on={value \thisrow{anchor}\as\myanchor}] 
	table [x expr=\thisrowno{0}, y expr=\thisrowno{1}/1, row sep=\\, meta=Label] {
x y Label anchor\\
1.93794 115.381 { Marlin(12,0) } {west} \\
1.98023 75.253 { Marlin(12,4) } {west} \\
2.02523 354.347 Rice {west} \\
0.996374 506.263 RLE {west} \\
1.24175 763.491 Snappy {west} \\
1.57133 2985.08 Nibble {west} \\
2.06672 314.004 FSE {west} \\
1.50688 316.592 Gipfeli {west} \\
1.88779 65.3695 Gzip {west} \\
1.25547 1164.06 Lzo {west} \\
2.06603 382.064 Huff0 {west} \\
1.19059 1055.64 Lz4 {west} \\
2.01858 197.722 Zstd {west} \\
2.11698 21.5154 CharLS {west} \\
	};

	\addplot[color=red,mark=*,only marks, point meta=explicit symbolic, nodes near coords,every node near coord/.append style={font={\sffamily\scriptsize},rotate=0, anchor=\myanchor},x=x,y=y,visualization depends on={value \thisrow{anchor}\as\myanchor}] 
	table [x expr=\thisrowno{0}, y expr=\thisrowno{1}/1, row sep=\\, meta=Label] {
x y Label anchor\\
2.02647 176.67 {Rice-Marlin(8,4)} {west}\\
	};
\end{axis}
\end{tikzpicture}

\begin{tikzpicture}
\begin{axis}[
x tick label style={font={\footnotesize},yshift=0mm}, 
y tick label style={font={\footnotesize},xshift=0mm},
xlabel style={font={\footnotesize},yshift= 2mm}, 
ylabel style={font={\footnotesize},yshift=-2mm},
height=5.5cm, width=14.5cm, ymin=0, ymax=4.25, xmin=1, xmax=2.5, 
xlabel=\emph{Compression Ratio}, ylabel=\emph{Decoding Speed (GiB/s)}]
	\addplot[color=black,mark=*,only marks, point meta=explicit symbolic, nodes near coords,every node near coord/.append style={font={\sffamily\scriptsize},rotate=0, anchor=\myanchor},x=x,y=y,visualization depends on={value \thisrow{anchor}\as\myanchor}] 
	table [x expr=\thisrowno{0}, y expr=\thisrowno{1}/1024, row sep=\\, meta=Label] {
x y Label anchor\\
1.93794 3181.64 { Marlin(12,0) } {south west} \\
1.98023 2626.68 { Marlin(12,4) } {south west} \\
2.02523 554.313 Rice {south east} \\
0.996374 521.419 RLE {south west} \\
1.24175 2230.39 Snappy {west} \\
1.57133 1036.25 Nibble {west} \\
2.06672 425.922 FSE {south west} \\
1.50688 368.344 Gipfeli {west} \\
1.88779 174.363 Gzip {east} \\
1.25547 2005.69 Lzo {west} \\
1.19059 3926.89 Lz4 {west} \\
2.01858 405.205 Zstd1 {east} \\
2.11698 24.4874 CharLS {south west} \\
	};

	\addplot[color=red,mark=*,only marks, point meta=explicit symbolic, nodes near coords,every node near coord/.append style={font={\sffamily\scriptsize},rotate=0, anchor=\myanchor},x=x,y=y,visualization depends on={value \thisrow{anchor}\as\myanchor}] 
	table [x expr=\thisrowno{0}, y expr=\thisrowno{1}/1024, row sep=\\, meta=Label] {
x y Label anchor\\
2.02647 2480.99 {Rice-Marlin(8,4)} {west}  \\
	};
\end{axis}
\end{tikzpicture}
\caption{
Evaluation on the Rawzor lossless image compression dataset~\cite{Rawzor}. 
Compared to Zstd, a very competitive fast compression algorithm, Rice-Marlin achieves comparable compression performance and encoding speed, while decoding 6.12x faster.
}
\label{fig:rawzor_image}
\end{figure}

In \figref{rawzor_image} we test Rice-Marlin against several state-of-the-art codecs on the Rawzor~\cite{Rawzor} image set compressing independently blocks of $64 \times 64$ pixels.
The prediction model uses the pixel above, and we compress the residuals.

Despite using dictionaries 16 times smaller, Rice-Marlin achieves a compression ratio of $2.02647$, a $2.34\%$ improvement over Marlin with Overlapping codes ~\cite{martinez2018improving}, and a $4.56\%$ improvement over original Marlin~\cite{martinez2017marlin}.
Despite the increased complexity of the new decoding algorithm, the decoding speed only drops by $5.56\%$.
On the other hand, Rice-Marlin is 2.4x faster to encode, achieving $176.67$ MiB/s, mainly thanks to the smaller dictionary, and the optimizations described in section~\ref{ch:implementation}.


\section{Conclusions}

We have presented two techniques for Marlin that improve the ratio between the dictionary size and the alphabet size by reducing the alphabet size. 
Using those we achieve better compression efficiency while using dictionaries 16 times smaller than before.
We presented an encoding algorithm that leverages the smaller dictionaries to achieve 2.4x times faster encoding times.
Thanks to this improvements, our new codec, named Rice-Marlin, achieves encoding speeds and compression ratios similar to the Zstd in lossless image compression, while still being 6.12x faster to decode.

\clearpage
\section*{References}

\bibliographystyle{IEEEtran}
\bibliography{dcc2019}

\begin{thebibliography}{10}
\providecommand{\url}[1]{#1}
\csname url@samestyle\endcsname
\providecommand{\newblock}{\relax}
\providecommand{\bibinfo}[2]{#2}
\providecommand{\BIBentrySTDinterwordspacing}{\spaceskip=0pt\relax}
\providecommand{\BIBentryALTinterwordstretchfactor}{4}
\providecommand{\BIBentryALTinterwordspacing}{\spaceskip=\fontdimen2\font plus
\BIBentryALTinterwordstretchfactor\fontdimen3\font minus
  \fontdimen4\font\relax}
\providecommand{\BIBforeignlanguage}[2]{{%
\expandafter\ifx\csname l@#1\endcsname\relax
\typeout{** WARNING: IEEEtran.bst: No hyphenation pattern has been}%
\typeout{** loaded for the language `#1'. Using the pattern for}%
\typeout{** the default language instead.}%
\else
\language=\csname l@#1\endcsname
\fi
#2}}
\providecommand{\BIBdecl}{\relax}
\BIBdecl

\bibitem{martinez2017marlin}
M.~Martinez, M.~Haurilet, R.~Stiefelhagen, and J.~Serra-Sagrist{\`a}, ``Marlin:
  A high throughput variable-to-fixed codec using plurally parsable
  dictionaries,'' in \emph{Proceedings of Data Compression Conference}.\hskip
  1em plus 0.5em minus 0.4em\relax IEEE, 2017.

\bibitem{martinez2018improving}
M.~Martinez, K.~Sandfort, D.~Dub{\'e}, and J.~Serra-Sagrist{\`a}, ``Improving
  marlin's compression ratio with partially overlapping codewords,'' in
  \emph{Proceedings of Data Compression Conference}.\hskip 1em plus 0.5em minus
  0.4em\relax IEEE, 2018.

\bibitem{LZ4}
Y.~Collet, ``{LZ}4,'' \url{lz4.github.io/lz4}, 2011.

\bibitem{LZO}
M.~Oberhumer, ``{LZO}: Lempel {Z}ip {O}berhumer,''
  \url{www.oberhumer.com/opensource/lzo}, 1996.

\bibitem{SNAPPY}
Z.~Tarantov and S.~Gunderson, ``Snappy,'' \url{google.github.io/snappy}, 2011.

\bibitem{deflate}
P.~Katz, ``Deflate lossless data compression,'' US patent 5051745, 1991.

\bibitem{ziv1977universal}
J.~Ziv and A.~Lempel, ``A universal algorithm for sequential data
  compression,'' \emph{IEEE Transactions on Information Theory}, vol.~23,
  no.~3, pp. 337--343, 1977.

\bibitem{huffman1952method}
D.~A. Huffman \emph{et~al.}, ``A method for the construction of
  minimum-redundancy codes,'' \emph{Proceedings of the IRE}, vol.~40, no.~9,
  pp. 1098--1101, 1952.

\bibitem{rissanen1976generalized}
J.~Rissanen, ``Generalized {K}raft inequality and arithmetic coding,''
  \emph{IBM Journal of Research and Development}, vol.~20, no.~3, pp. 198--203,
  1976.

\bibitem{martin1979range}
G.~N.~N. Martin, ``Range encoding: an algorithm for removing redundancy from a
  digitised message,'' in \emph{Proc. Institution of Electronic and Radio
  Engineers International Conference on Video and Data Recording}, 1979.

\bibitem{lenhardt2012gipfeli}
R.~Lenhardt and J.~Alakuijala, ``Gipfeli-high speed compression algorithm,'' in
  \emph{Proceedings of Data Compression Conference}.\hskip 1em plus 0.5em minus
  0.4em\relax IEEE, 2012, pp. 109--118.

\bibitem{fastLZ}
R.~N. Williams, ``An extremely fast {Z}iv-{L}empel data compression
  algorithm,'' in \emph{Proceedings of Data Compression Conference}.\hskip 1em
  plus 0.5em minus 0.4em\relax IEEE, 1991, pp. 362--371.

\bibitem{harnik2014fast}
D.~Harnik, E.~Khaitzin, D.~Sotnikov, and S.~Taharlev, ``A fast implementation
  of {D}eflate,'' in \emph{Proceedings of Data Compression Conference}.\hskip
  1em plus 0.5em minus 0.4em\relax IEEE, 2014, pp. 223--232.

\bibitem{Huff0}
Y.~Collet, ``Huff0,''
  \url{fastcompression.blogspot.de/p/huff0-range0-entropy-coders.html}, 2013.

\bibitem{FSE}
------, ``Finitestateentropy,'' \url{github.com/Cyan4973/FiniteStateEntropy},
  2013.

\bibitem{DBLP:journals/corr/abs-0902-0271}
\BIBentryALTinterwordspacing
J.~Duda, ``Asymmetric numeral systems,'' \emph{CoRR}, vol. abs/0902.0271, 2009.
  [Online]. Available: \url{http://arxiv.org/abs/0902.0271}
\BIBentrySTDinterwordspacing

\bibitem{Rice}
R.~Rice and J.~Plaunt, ``Adaptive variable-length coding for efficient
  compression of spacecraft television data,'' \emph{IEEE Transactions on
  Communication Technology}, vol.~19, no.~6, pp. 889--897, 1971.

\bibitem{Lagarith}
B.~Greenwood, ``Lagarith,'' \url{lags.leetcode.net/codec.html}, 2011.

\bibitem{CharLS}
J.~de~Vaan, ``{CharLS},'' \url{github.com/team-charls/charls}, 2007.

\bibitem{tunstall1967synthesis}
B.~P. Tunstall, ``Synthesis of noiseless compression codes,'' \emph{Ph.D.
  dissertation, Georgia Institute of Technology}, 1967.

\bibitem{savari1999variable}
S.~A. Savari, ``Variable-to-fixed length codes and plurally parsable
  dictionaries,'' in \emph{Proceedings of Data Compression Conference}.\hskip
  1em plus 0.5em minus 0.4em\relax IEEE, 1999, pp. 453--462.

\bibitem{al2015using}
A.~Al-Rababa'a and D.~Dub{\'e}, ``Using bit recycling to reduce the redundancy
  in plurally parsable dictionaries,'' in \emph{14th Canadian Workshop on
  Information Theory}.\hskip 1em plus 0.5em minus 0.4em\relax IEEE, 2015, pp.
  62--65.

\bibitem{yoshida2010efficient}
S.~Yoshida and T.~Kida, ``An efficient algorithm for almost instantaneous {VF}
  code using multiplexed parse tree,'' in \emph{Proceedings of Data Compression
  Conference}.\hskip 1em plus 0.5em minus 0.4em\relax IEEE, 2010, pp. 219--228.

\bibitem{yamamoto2001average}
H.~Yamamoto and H.~Yokoo, ``Average-sense optimality and competitive optimality
  for almost instantaneous {VF} codes,'' \emph{IEEE Transactions on Information
  Theory}, vol.~47, no.~6, pp. 2174--2184, 2001.

\bibitem{dube2018individually}
D.~Dub{\'{e}} and F.~Haddad, ``Individually optimal single- and multiple-tree
  almost instantaneous variable-to-fixed codes,'' in \emph{2018 {IEEE}
  International Symposium on Information Theory}, 2018, pp. 2192--2196.

\bibitem{Rawzor}
S.~Garg, ``The new test images,'' \url{www.imagecompression.info/test_images},
  2011.

\end{thebibliography}

 \end{document}